\documentstyle[aps,pra,multicol,floats,epsfig]{revtex}
\newcommand{\beq}{\begin{equation}}
\newcommand{\enq}{\end{equation}}
\newcommand{\tpnew}{\bf}

\begin{document}

\draft


\title{Generation and evolution of vortex-antivortex pairs in
Bose-Einstein condensates}

\author{J.-P. Martikainen$^1$, K.-A. Suominen$^{1,2}$,
L. Santos$^3$, T. Schulte$^3$, and A. Sanpera$^3$}

\address{$^1$Helsinki Institute of Physics, PL 64, FIN-00014
Helsingin yliopisto, Finland\\
$^2$Department of Applied Physics, University of Turku, FIN-20014,
Turun yliopisto, Finland\\
$^3$Institut f\"{u}r Theoretische Physik, Universit\"{a}t Hannover,
30167 Hannover, Germany
}

\date{\today}

\maketitle

\begin{abstract}

We propose a method for generating and controlling a spatially
separated 
vortex--antivortex pair in a
Bose-Einstein condensate trapped in a toroidal potential. 
Our simulations of the time dependent Gross-Pitaevskii equation
show that in toroidal condensates vortex dynamics are 
different from the dynamics in  the homogeneous case. 
Our numerical results agree well with 
analytical calculations using the image method. 
Our proposal offers an
effective example of coherent generation and control of vortex dynamics
in atomic condensates.
\end{abstract}

\pacs{03.75.Fi, 32.80.Pj, 03.75.-b} 

\begin{multicols}{2}
\narrowtext

\section{Introduction}
Quantum liquids and gases are an excellent environment for studies of
solitons and topological quantum structures such as vortices. Studies of
superfluidity and vorticity have been done mostly in
He\,\rm{II}~\cite{Donnelly91}. Due to the high density and strong
interactions the theoretical studies give usually only a qualitative
agreement with experiments. Moreover, the size of the vortex core is very
small in He\,\rm{II}, typically $\sim 1$~\AA. The situation is
drastically different for atomic Bose-Einstein
condensates~\cite{FirstBEC}. These many-body systems are dilute and
weakly interacting, and as a result a quantitative agreement between
theory and experiments can often be found. Also, the expected size of a
vortex core is fairly large, {\tpnew $\sim 0.1-1\,\mu\rm{m}$}, making
detailed experimental studies possible.

Several theoretical approaches have been suggested for the
generation of vortices in condensates, see
e.g.~Refs.~\cite{Martzlin97,Dum98,Jackson98,Caradoc99,Dobrek99,Williams99,Feder99,Feder00}. 
Among them, the stirring of a condensate with a blue-detuned
laser in a single component~\cite{Madison20}, and the coherent
interconversion  between the two components of a binary
condensate~\cite{Matthews99} have been successfully implemented. In
addition, the creation of vortex rings after the decay of a dark soliton 
has been recently reported~\cite{Anderson00}. These vortices carry one quantum of
circulation each, and they form stable, geometric patterns. However, the
success in controlling atomic  condensates suggests that one could create
also non-equilibrium situations to investigate dynamics of vortices, and
in particular, vortex collisions. Alternatively, a
vortex can be generated through the ``phase-imprinting''
method~\cite{Dobrek99}.  This approach has been successfully used
to generate solitons in condensates~\cite{Burger99}.


Here
we propose how to create and control a pair of spatially
separated vortices of opposite circulation via controlled decay of
dynamically unstable solitons in a toroidal trapped condensate.
Due to the toroidal geometry and the employed technique for the creation
of the vortices, the dynamics and circulation of the generated vortices
are correlated, even when they are generated at well-separated locations.

In homogeneous condensates and in absence of dissipation
a vortex--antivortex pair moves as a whole through the
fluid~\cite{Donnelly91}. Such behavior can be, however, substantially
modified in the case of trapped condensates. In particular, for the case
of stiff potentials the vortex dynamics is altered by the constraint of
vanishing normal velocity field at the boundaries \cite{Fetter67}.
We show how vortex--antivortex dynamics in toroidal
geometries become 
different from the equivalent dynamics
in a homogeneous (not trapped) condensate.



We propose a combination of condensate splitting, phase
imprinting, and subsequent merging of separated parts into a toroidal
condensate, which leads to a very controlled mechanism of vortex creation.
We describe the details of our proposal and numerical solution of the relevant 
Gross-Pitaevskii equation in Sec. \ref{Numerics}. In Sec. \ref{Images}
we apply the method of images to explain the vortex dynamics and we
give some concluding remarks in Sec. \ref{Concl}

\section{Vortex creation and dynamics numerically}\label{Numerics}
We assume initially a condensate trapped in a Gaussian optical trap. By
shaking the trap rapidly we split the condensate into two spatially
separated parts~\cite{Dum98b}. First we slowly increase the amplitude of
the periodic shaking along the chosen $x$-axis~\cite{Shaking}. The
effective potential evolves smoothly from the Gaussian well into a trap
with two separated minima; this is the time-averaged potential.
The condensate follows adiabatically the changes of the time-averaged
potential, splitting into two parts (Fig.~\ref{fig1}). By adding a similar
process along the $y$-axis the time-averaged trapping potential evolves
towards a torus. Thus the two separated condensate parts combine
eventually into a torus. Before merging the two halves, we imprint on one
of them (with a strongly detuned laser field~\cite{Dobrek99}) a change of
phase of the wavefunction, $\Delta \phi$.

At sufficiently low temperatures, the system is well described by the 
corresponding Gross-Pitaevskii equation
\beq
   i\hbar\frac{\partial\Psi}{\partial t}=
   -\frac{\hbar^2}{2m}\nabla^2\Psi+V\Psi+NU_0|\Psi|^2\Psi.
\label{GP}
\enq
Here $\Psi$ is the condensate wavefunction, $V(x,y,t)$ is the trap
potential~\cite{Shaking} (not the time-averaged one), $N$ is the number of
atoms and $U_0=4\pi\hbar^2a/m$, where $a$ is the $s$-wave scattering
length and $m$ is the atomic mass.  In all simulations we have assumed a
sodium condensate ($a=2.75$ nm) with the particle number $N=10^5$.
The width of the trap in $x$- and $y$-directions was $10\ \mu$m and its
size in the $z$-direction $L_z=2\ \mu$m. The trap depth was set to
$4\,\mu\rm{K}$.  In our simulations the duration of each shaking stage
was typically 50 ms. The creation of the torus can be seen in the first
steps of Fig.~\ref{fig1}. With the chosen parameters the trap ground state
energy in $z$-direction is considerably larger than the mean field
energy $nU_0$ at the toroidal trap ($n$ is the atomic density). This fact allows us 
to restrict ourselves to a two--dimensional geometry.


The coherence time for the applied phase shift depends on the details of
the shaking. If we express the wavefunction as $\Psi=R\exp(i\phi)$ and
assume the Thomas-Fermi limit, i.e., ignore terms due to kinetic energy in
Eq.~(\ref{GP}), we obtain (assuming that $R$ changes sufficiently slowly)
\beq
    \label{phase_eq}
	  \frac{\partial\phi}{\partial t}=-\frac{V(r,t)+NU_0R(r,t)^2}{\hbar}.
\enq
Since in the Thomas-Fermi limit the right-hand side of Eq.~(\ref{phase_eq})
becomes independent of $r$, the phase evolution is identical everywhere.
During the time between applying the phase shift and the merging into a
torus, the phase has changed by a constant amount over the condensate
halves and thus there will be a  discontinuity in the phase across the
merging points, corresponding to the applied phase.
If, on the other hand, the condensate exhibits sloshing, the kinetic
energy term makes the phase evolution $r$-dependent and the applied phase
shift loses its meaning. Even a small asymmetry between the two parts is
then enough to mask the applied phase shift on the time-scale of $\sim
100$ ms, which in effect leads to a total loss of the control of the
vortex production. In our simulations the density distribution followed
the Thomas-Fermi limit closely and, therefore, the phase evolution was
almost identical everywhere.

When the phase-shifted condensate parts form a torus, the discontinuity
in phase is quickly transformed into a dark soliton, since the system
can adapt to the phase discontinuity only by reducing its density.
In general, such a kink-like state in a trapped condensate is dynamically
stable if the mean-field energy $nU_0$ is roughly smaller than the
characteristic  trap energy in the direction given by the soliton
front~\cite{Muryshev99}, which in our case corresponds to the typical energy of 
the trap in the radial direction. 
With the parameters from our setup \cite{Shaking}
and $n\sim 1.7\,\cdot 10^{14}$ cm$^{-3}$,  this
condition is violated. Under such conditions the soliton front undergoes
snake instabilities  which can eventually decay into vortex-antivortex
pairs  (or vortex rings in 3D configurations),
see
e.g.~\cite{Feder00,Muryshev99,Kuznetsov88,Josserand95,Kivshar98,Brand01}.  
This process has been experimentally
observed in non-linear optics~\cite{Tikhonenko96}, as well as, very
recently, in matter waves~\cite{Anderson00}. Since the front perturbation
with a wavevector
$k=1/(\sqrt{2}\xi)$, where
$\xi=1/\sqrt{4\pi na}$ is the healing length,  will grow
fastest~\cite{Muryshev99}, the distance between the generated vortices is
expected to be about $d\sim\pi/k$. If $k>1/\xi$ the normal modes are
dynamically stable, i.e., if the torus is narrower than about $\pi\xi$
the soliton is dynamically stable.

The depth of the soliton depends on the size of the phase discontinuity.
It must be large enough for vortex generation. Otherwise the resulting
soliton can  decay into other excitations, or even remain stable.  For
example, a $\pi/2$ phase shift did not lead to generation of vortices in
our simulations. Due to the $2\pi$ degeneracy of the phase, the phase
shift $\pi$ corresponds to the largest possible discontinuity. But one
should note that the direction of the phase gradient over the
discontinuity affects the subsequent evolution. If the phase difference
is equal to $\pi$, the $2\pi$ degeneracy makes the direction of the
gradient ambiguous. In our simulations the motion of vortices was
initially indeterminate and thus very sensitive to small disturbances,
when $\Delta\phi\simeq\pi$.  For $\Delta\phi=0.9\pi$ (as in
Fig.~\ref{fig1}), however, we obtain a much more predictable behavior.
Due to symmetry, the mirror image of this situation is obtained with
$\Delta\phi= 1.1\pi$ (the lower vortex moves clockwise in the beginning,
and the upper one counterclockwise in Fig.~\ref{fig1}).

In our simulations each soliton decays into a pair of vortices of
opposite circulation. However, one of the vortices, located close to
the edge of the condensate,  has a very short lifetime, and decays
within a few ms at the edge. The survival of only one vortex per soliton
is not surprising, considering that $d\sim l$, where $l$ is the
width of the torus. For wider torii, i. e., longer solitons, our
simulations show a decay of each soliton into many vortex-antivortex
pairs, as expected~\cite{Josserand95}. Similar behaviour has been reported
in simulations of condensates in rectangular boxes~\cite{Carr00}.  In our
situation the condensate wave function  remains symmetric with respect to
reflection to the horizontal axis (see Fig.~\ref{fig1}).  This symmetry
leads to 
correlations between vortex production at the two
phase discontinuities, located on opposite sides of the torus. We are
always left with two vortices (one vortex per soliton) of opposite
circulations. These vortices have also opposite vertical and equal
horizontal components of the velocity, and they are therefore
automatically set on a collision course. This reflects the conservation of
angular momentum (we have verified that the shaking process or the
adiabatic merging of the separated condensate parts do not bring any
extra angular momentum into the system)


Figure~\ref{fig1} shows the time evolution obtained from numerical
simulations. The created vortices bounce from each other when they
approach.  This process repeats itself if we continue simulations beyond
$t=300$ ms. In a nondissipative homogeneous system two vortices of
opposite circulation  separated by a distance greater than the healing
length move parallel, since each vortex will move with the velocity of
the other one\cite{Donnelly91}.  
Clearly, the vortex dynamics in a torus
trap is more complex and completely different. It can be understood
with the help of the images method. 

\section{Images method}\label{Images}
   The images method has been recently
successfully employed to describe both a single vortex\cite{Fedichev99},
and vortex arrays dynamics in circular 2D traps~\cite{graham}.
We consider in the following a stiff torus (with infinitely
high and stiff walls), i.e. we neglect the influence of the inhomogeneous trapping potential on the vortex dynamics~\cite{Rubinstein94,Svidzinsky00}.
Additionally, we assume that the distance between the
vortices is greater than the size of the vortex cores. 
In absence of friction the vortex velocity equals the superfluid velocity at the vortex position. Such a velocity field, 
which is induced by the presence of the vortices, has to fulfill the
constraint of vanishing normal component at the torus boundaries.
To fulfill this constraint it is enough to consider for each vortex an infinite 
number of fictitious image vortices, each with appropriate position and circulation.
Each vortex, either real or image, contributes to the superfluid velocity field ${\bf v}_{SF}({\bf r})$ by
$\left(\kappa_i/2\pi\right)\hat{\bf z}
\times({\bf r}-{\bf r}_i)/|{\bf r}-{\bf r}_i|^2$,
where ${\bf r}_i$ and $\kappa_i$ refer to its position and 
circulation, respectively.  

Let us consider the case of a torus 
with an inner radius $R_{1}$, an outer radius $R_{2}$,
and one vortex with circulation $\kappa$ 
located at ${\bf r}$ (in a frame centered at the origin of the torus).
In order to cancel the normal component of the superfluid velocity on the boundaries,
we have to consider two different families of image vortices. For the first family, an image vortex is placed at ${\bf r}_{im1}=\frac{R^{2}_{1}}{r}\frac{{\bf r}}{r}$
with circulation $-\kappa$ as well as one image vortex with circulation $\kappa$ at the origin~\cite{Saffman92}.
The image vortex at ${\bf r}_{im1}$ will induce a velocity component
normal on the outer circle, so in addition we need an image vortex at
${\bf r}_{im2}=\frac{R^{2}_{2}}{r_{im1}} \frac{{\bf r}}{r} = \frac{{R^{2}_{2}}}{R^{2}_{1}} \bf r$
with circulation $\kappa$.
In turn this image will induce some normal
component on the inner circle, so two more image vortices are required: one at ${\bf r}_{im3}=
\frac{R^{2}_{1}}{r_{im2}} \frac{{\bf r}}{r} = \frac{R^{4}_{1}}{R^{2}_{2}\, r} \frac {\bf r}{r}$ with circulation $-\kappa$ the other one at the origin with circulation $ \kappa$  
and so on and so forth. 
For the second family, one considers an image vortex at 
${\bf r'}_{im1}=\frac {R^{2}_{2}}{r} \frac {\bf {r}}{r}$ with circulation $-\kappa$.
This image vortex gives a normal component of the velocity on the inner
circle so we need in addition one image vortex at
${\bf r'}_{im2}=\frac {R^{2}_{1}}{r'_{im1}} \frac {\bf r}{r}= \frac {R^{2}_{1}}{R^{2}_{2}}  {\bf r}$
with circulation $\kappa$ as well as one image vortex at the origin with circulation $- \kappa$. For this we require another image vortex at $ {\bf r'}_{im3}=\frac {R^{2}_{2}}{r'_{im2}} \frac {\bf r}{r}= \frac {R^{4}_{2}}{R^{2}_{1} \,r} \frac {\bf r}{r}$
with circulation $-\kappa$ and so on.
For the second real vortex in the torus we have to repeat the same formalism to get an image configuration, which ensures that the boundary condition is fulfilled.

That way we can express the total superfluid velocity field in the torus (for arbitrary circulations), in particular the velocity at the locations of the real vortices as a function of their mutual positions. Since the latter equals the vortex velocity we get the following first order equations for the vortex positions ${\bf r}_i$ that can be easily computed 

\end{multicols}
\widetext

\begin{eqnarray}
2\pi \frac{d}{dt}\, \overrightarrow{r}_{i} & = & \sum ^{2}_{j=1}\sum ^{\infty }_{n=1}\: \left( \: \frac{\kappa _{j}\: \left( r_{i}\, \overrightarrow{e}_{\varphi _{i}\: -}\: r_{j}\, \left( \frac{R_{2}}{R_{1}}\right) ^{2n}\: \overrightarrow{e}_{\varphi _{j}}\right) }{\left( \overrightarrow{r}_{i}-\left( \frac{R_{2}}{R_{1}}\right) ^{2n}\overrightarrow{r}_{j}\right) ^{2}}\quad +\quad R_{1}\: \leftrightarrow \: R_{2}\;\, \right) \nonumber \\
 & - & \sum ^{2}_{j=1}\sum ^{\infty }_{n=0}\: \left( \: \frac{\kappa _{j}\: \left( r_{i}\, \overrightarrow{e}_{\varphi _{i}\: -}\: \left( \frac{R_{1}}{R_{2}}\right) ^{2n}\: \frac{R^{2}_{1}}{r_{j}}\: \overrightarrow{e}_{\varphi _{j}}\right) }{\left( \overrightarrow{r}_{i}-\left( \frac{R_{1}}{R_{2}}\right) ^{2n}\frac{R^{2}_{1}}{r^{2}_{j}}\overrightarrow{r}_{j}\right) ^{2}}\quad +\quad R_{1}\: \leftrightarrow \: R_{2}\: \right) \nonumber \\
 & + & \; \frac{\kappa _{j\neq i}\: \left( r_{i}\, \overrightarrow{e}_{\varphi _{i}\:}-\: r_{j\neq i}\: \overrightarrow{e}_{\varphi _{j\neq i}}\right) }{\left( \overrightarrow{r}_{i}-\overrightarrow{r}_{j\neq i}\right) ^{2}} \quad + \quad \sum ^{2}_{j=1}\frac{\kappa _{j}}{r_{i}}\: \overrightarrow{e}_{\varphi _{i}}\qquad \qquad \qquad \qquad \qquad i,j=1,2
\end{eqnarray}

\begin{multicols}{2}

\noindent where $ R_{1}\: \leftrightarrow \: R_{2}$ 
indicates a term equal to the previous one but interchanging $R_1$ and $R_2$.
The results obtained with the image method are presented in
Fig.~\ref{fig2}.  As initial conditions we take two vortices located on
opposite parts of the torus and displaced symmetrically from the central
point between the torus boundaries. For the sake of clarity our initial
conditions have been chosen such that the vortex trajectories are wider,
i.e., they explore more space between the  torus boundaries than in the
numerical simulation from Fig.~\ref{fig1}. But with appropriate initial
vortex positions one can achieve a nearly perfect agreement with
simulations of Eq.~(\ref{GP}) concerning the shape of the trajectories,
and the period of their oscillations ($\simeq 300$ ms).

\section{Conclusions}\label{Concl}
So far, we have assumed that the condensate is at zero temperature, i.e.,
we have used Gross-Pitaevskii equation~(\ref{GP}) to model the condensate,
neglecting any dissipation. The presence of non-condensate atoms  is
expected to result in dissipative Magnus forces and vortex decay. But
experimental studies~\cite{Madison20} so far show that a realistic time
scale for vortex decay is a few seconds, which is longer than the
time scale of our scenario. We tested the stability of our system by
adding a small, constant imaginary term into Eq.~(\ref{GP}). 
It did not
modify the reported behavior unless the magnitude of the term was
too strong to be considered a perturbation.
 Nevertheless, the problem of the
effects of dissipation in our model is very challenging. The
presence of dissipation might cause the vortex-antivortex pair
either to collide with the trapping boundaries, or to mutually annihilate.

Another complication arises from the fact that currently the spatial resolution of the
measurement of the condensate density profile is limited. Usually this problem
is avoided by letting the condensate  expand rapidly by removing the
trapping potential. In a quasi-2D traps the optical depth will be an issue.
Therefore some other detection scheme, such as matter wave 
interference~\cite{Dobrek99,Bolda1998a,Tempere1998c,Castin1999a,Chevy2001a},  
may have to be considered.




Summarizing, we have presented a method for creating vortices in a toroidal geometry 
in a controllable way. This method allows for the analysis of the interaction of vortices 
in such a geometry, which is significantly modified by the geometry of the trapping potential. 
The realm of possible phenomena that can be generated and controlled by
merging the split condensate parts into a torus is not by any means
limited by those discussed here, e.g. if we form the torus rapidly, 
other excitations are also created, which
interact with the vortices and affect their dynamics.

In the analytical estimation of the vortex orbits we have neglected for simplicity the effects of the 
inhomogeneity of the trapping potential. Such inhomogeneity will result in an additional contribution to 
the velocity of the vortex in the direction perpendicular to the density gradient, as discussed in Ref.\ 
\cite{Svidzinsky00} (a detailed analysis will be the subject of a later publication).
On the other hand, our calculations have been constrained to quasi-2D traps. Very recently the issue of lower dimensional BEC has been subject of great interest. In particular 2D (and even 1D) condensates have been experimentally observed~\cite{Safonov98,Görlitz}. Therefore the situation considered in the present paper is experimentally justified and feasible. However, the analysis of 3D geometries constitutes a challenging problem. We have made some short simulations to study the dynamics of vortex-antivortex (line) interactions in a 3D pipe geometry. The results we obtain are very similar to those presented here. In a proper 3D geometry, the solitons decay into vortex rings, as observed very recently in the case of cylindrical traps~\cite{Anderson00}. The propagation, deformation and interaction of vortex rings will be the topic for future studies.

We thank M. Lewenstein, G. Shlyapnikov and M. Baranov for discussions.
We acknowledge support of the Academy of Finland (project  43336), Center
for Scientific Computing (CSC), and Deutsche Forschungsgemeinschaft (SFB
407). J.-P. M. is supported by the National Graduate School on Modern
Optics and Photonics.

\end{multicols}

\begin{figure}[h]
\vspace*{15cm}
\vspace*{1mm}
\caption[fig1]{Numerical simulation of the creation and dynamics of a
vortex-antivortex pair in a Bose-Einstein condensate. The  condensate is
split into two parts, and then a phase shift $\Delta\phi=0.9\pi$ is
applied to the  left part of the wavefunction. As the two parts merge to
form a torus two solitons which subsequently decay into vortices are
created. The two surviving vortices move along the torus and bounce from
each other as they collide (minimum distance at $t=186$ ms).
\label{fig1}}
\end{figure}

\begin{figure}[hbt]
\centerline{\psfig{figure=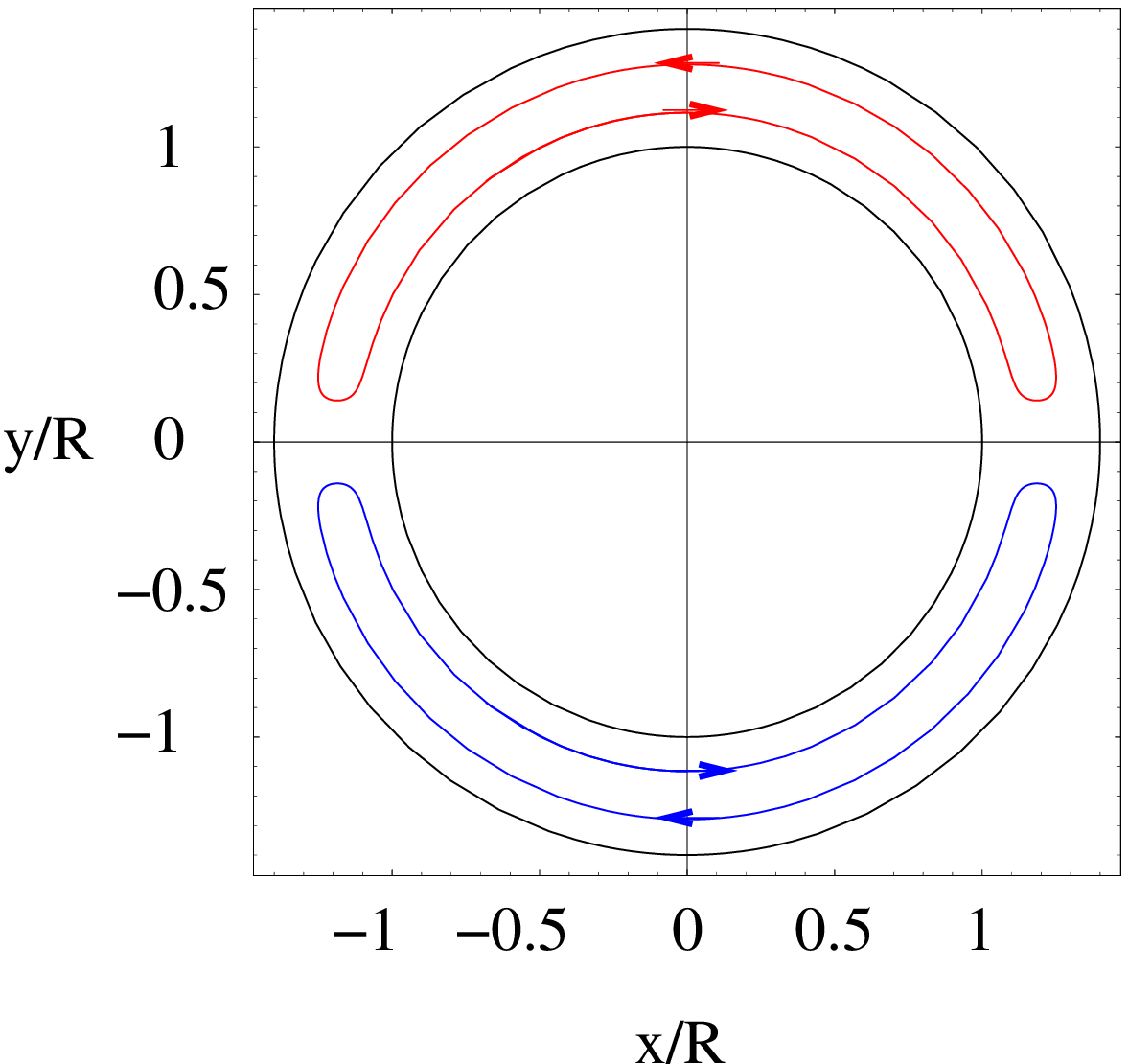,width=75mm}}
\caption[f2]{
Description of the vortex-antivortex dynamics in a stiff torus
using the image method. The trajectories of the vortex and antivortex are
shown.}
\label{fig2}
\end{figure}


\begin{references}
\bibitem{Shaking} The trap potential is
$-V_0\exp[-(x-x_0)^2/(2\sigma_x^2)-(y-y_0)^2/(2\sigma_y^2)]$, where
$V_0/k_B=4\ \mu$K, $\sigma_x=\sigma_y=10\ \mu$m. We shake it by setting
$x_0=r_0\sin(\pi t/T)\cos(\omega_rt)$ and $y_0=0$ for $0<t<T/2$,
$x_0=r_0\cos(\omega_rt)$ and $y_0=r_0\sin[\pi (t-T/2)/T]\sin(\omega_rt)$
for $T/2<t<T$, and $x_0=r_0\cos(\omega_rt)$ and $y_0=r_0\sin(\omega_rt)$
for $t>T$. Here $r_0=20\ \mu$m, $\omega_r=(2\pi) 6.6$ kHz, $T=100$
ms.
\end{references}
\end{document}